\documentclass[12pt,a4paper]{article}\usepackage[]{graphicx}\usepackage{xcolor}
\makeatletter
\def\maxwidth{ %
  \ifdim\Gin@nat@width>\linewidth
    \linewidth
  \else
    \Gin@nat@width
  \fi
}
\makeatother

\definecolor{fgcolor}{rgb}{0.251, 0.251, 0.282}

\usepackage{framed}
\makeatletter
 {\par\unskip\endMakeFramed%
 \at@end@of@kframe}
\makeatother

\definecolor{shadecolor}{rgb}{.97, .97, .97}
\definecolor{messagecolor}{rgb}{0, 0, 0}
\definecolor{warningcolor}{rgb}{1, 0, 1}
\definecolor{errorcolor}{rgb}{1, 0, 0}

\usepackage{alltt}
\usepackage[margin=1.3in]{geometry}

\usepackage[]{graphicx}
\usepackage{xcolor}

\usepackage{alltt}

\usepackage{bm,amsmath,amsthm,amsfonts,amssymb}
\usepackage{natbib}

\theoremstyle{plain}
\theoremstyle{remark}

\newtheorem{res}{Result}
\usepackage{algorithm}
\usepackage{algpseudocode}
\usepackage{tikz}

\usepackage[colorlinks = true,linkcolor = blue, citecolor = blue]{hyperref}

\newcommand{\myalphafoot}
{
\renewcommand{\thefootnote}{\alph{footnote}}
}

\title{Sequential Unequal Probability Sampling For Stream Population}
\myalphafoot
\author{\myalphafoot  Bardia Panahbehagh\footnotemark[1]~, Rapha\"el Jauslin\footnotemark[2]~ and Yves Till\'e\footnotemark[2]}
\date{}
\footnotetext[1]{Department of Mathematics, Kharazmi University, Tehran, Iran (E-mail: bardia.panahbehagh@unine.ch)}
\footnotetext[2]{Institute of statistics, University of Neuch\^atel, Bellevaux 51, 2000 Neuch\^atel, Switzerland}

\def\vva{$\scriptstyle\begin{pmatrix}0.5\\	0.5	\\	0.3	\\	0.1	\\	0.6	\\	0.7	\\	0.3\end{pmatrix}$}
\def\vvb{$\scriptstyle\begin{pmatrix}0	\\	1	\\	0.3	\\	0.1	\\	0.6	\\	0.7	\\	0.3\end{pmatrix}$}
\def\vvc{$\scriptstyle\begin{pmatrix}1	\\	0	\\	0.3	\\	0.1	\\	0.6	\\	0.7	\\	0.3\end{pmatrix}$}
\def\vvd{$\scriptstyle\begin{pmatrix}0	\\	1	\\	0	\\	0.14\\	0.86\\	0.7	\\	0.3\end{pmatrix}$}
\def\vve{$\scriptstyle\begin{pmatrix}0	\\	1	\\	1	\\	0	\\	0	\\	0.7	\\	0.3\end{pmatrix}$}
\def\vvf{$\scriptstyle\begin{pmatrix}1	\\	0	\\	0	\\	0.14\\	0.86\\	0.7	\\	0.3\end{pmatrix}$}
\def\vvg{$\scriptstyle\begin{pmatrix}1	\\	0	\\	1	\\	0	\\	0	\\	0.7	\\	0.3\end{pmatrix}$}
\def\vvh{$\scriptstyle\begin{pmatrix}0	\\	1	\\	0	\\	0	\\	1	\\	0.7	\\	0.3\end{pmatrix}$}
\def\vvi{$\scriptstyle\begin{pmatrix}0	\\	1	\\	0	\\	1	\\	0	\\	0.7	\\	0.3\end{pmatrix}$}
\def\vvj{$\scriptstyle\begin{pmatrix}0	\\	1	\\	1	\\	0	\\	0	\\	0	\\	1\end{pmatrix}$}
\def\vvk{$\scriptstyle\begin{pmatrix}0	\\	1	\\	1	\\	0	\\	0	\\	1	\\	0\end{pmatrix}$}
\def\vvl{$\scriptstyle\begin{pmatrix}1	\\	0	\\	0	\\	0	\\	1	\\	0.7	\\	0.3\end{pmatrix}$}
\def\vvm{$\scriptstyle\begin{pmatrix}1	\\	0	\\	0	\\	1	\\	0	\\	0.7	\\	0.3\end{pmatrix}$}
\def\vvn{$\scriptstyle\begin{pmatrix}1	\\	0	\\	1	\\	0	\\	0	\\	0	\\	1\end{pmatrix}$}
\def\vvo{$\scriptstyle\begin{pmatrix}1	\\	0	\\	1	\\	0	\\	0	\\	1	\\	0\end{pmatrix}$}
\def\vvp{$\scriptstyle\begin{pmatrix}0	\\	1	\\	0	\\	0	\\	1	\\	0	\\	1\end{pmatrix}$}
\def\vvq{$\scriptstyle\begin{pmatrix}0	\\	1	\\	0	\\	0	\\	1	\\	1	\\	0\end{pmatrix}$}
\def\vvr{$\scriptstyle\begin{pmatrix}0	\\	1	\\	0	\\	1	\\	0	\\	0	\\	1\end{pmatrix}$}
\def\vvs{$\scriptstyle\begin{pmatrix}0	\\	1	\\	0	\\	1	\\	0	\\	1	\\	0\end{pmatrix}$}
\def\vvt{$\scriptstyle\begin{pmatrix}1	\\	0	\\	0	\\	0	\\	1	\\	0	\\	1\end{pmatrix}$}
\def\vvu{$\scriptstyle\begin{pmatrix}1	\\	0	\\	0	\\	0	\\	1	\\	1	\\	0\end{pmatrix}$}
\def\vvv{$\scriptstyle\begin{pmatrix}1	\\	0	\\	0	\\	1	\\	0	\\	0	\\	1\end{pmatrix}$}
\def\vvw{$\scriptstyle\begin{pmatrix} 1	\\	0	\\	0	\\	1	\\	0	\\	1	\\	0\end{pmatrix}$}
\IfFileExists{upquote.sty}{\usepackage{upquote}}{}

\IfFileExists{upquote.sty}{\usepackage{upquote}}{}
\begin{document}
	
	\maketitle

	
	\begin{abstract}
		A new unequal probability sampling method is proposed. This method is sequential. The decision to select or not each unit is made based on the order in which the units appear. A variant of this method allows to select a sample from a stream. At each step, the decision to take the units successively according to the order of appearance in the stream is made. This method involves using a sliding window that is as small as possible. The method also allows the sample to be spread and even the level of spreading to be adjusted.
	\end{abstract}
	
	\clearpage
	\section{Introduction}\label{intro}
	
	There are dozens of methods of unequal probability sampling. Most of them are described in the following books~\cite{han:bre:80, gab:90, til:06}. Nevertheless, very few methods allow you to decide to select the units in the order in which they appear.
	Systematic sampling with unequal probabilities~\citep{mad:49} allows sequential sampling from a stream. However, once the first units are examined, this method quickly becomes deterministic, making it predictable, which can be problematic. Indeed, sampling cannot be predictable if it is used to perform a control~\citep{bust:til:2020}.
	
	The ordered pivotal method~\citep{dev:til:98} is an alternative to systematic sampling. It is a special case where pairs of units are processed according to their order in the data list. Pivotal method and Deville's systematic sampling~\citep{dev:98a,til:06} are two different imple\-mentations of the same sampling design~\citep{cha:12}. A disadvantage of the pivotal method is that the decision to take or not take the first unit may be made after examining a very large number of units in the sampling frame.
	
	There are also so-called reservoir methods that allow sampling from a stream. Reservoir's size could be the sample size~\citep{cha:82}. It is updated each time a unit enters in the stream~\citep[see also][]{cohen2009stream}. Reservoir methods were generalized to select balanced samples~\citep{til:19a}. However, all these methods suffer from the same problem: the decision about a unit that is at the beginning of the stream can be made very late.
	
	The proposed method allows us to make decisions about the selection of units according to their order of appearance in a list. The update of the probabilities is realized only on a window containing a very small number of units which follows the unit for which the decision is made. The method is therefore particularly interesting for selecting units from a stream. The great advantage of this new method is that it forces the decision on the units according to their order of appearance in the stream.
	
	Section~\ref{Nota} introduces the notations and the principal concept of sampling theory. In Section~\ref{MainD}, we present the proposed sampling method while Section~\ref{streams} extends the method to stream sampling. Section~\ref{sec:spread} and Section \ref{sec:nonint} discusses different limit cases of the method and how the problems are solved. Section~\ref{sec:simu} exemplifies the method on a real data set and shows that the method is comple-tely comparable in terms of variance. The manuscript ends with a conclusion on the proposed method.
	
	
	\section{Notation}\label{Nota}
	
	Consider a finite population of size $N$ labeled by $U=\{1,2,\dots,N\}$, a vector of values taken by a variable of interest $\bm{y}=(y_1,y_2,\dots ,y_N)^\top$ and a vector of first-order inclusion probabilities $\bm{\pi}=(\pi_1,\pi_2,\dots,\pi_N)^\top$, each associated with the respective labels in $U$. Consider also the total of the variable of interest,
	$$
	Y=\sum_{k\in U} y_k,
	$$
	as the main parameter. A sample $s$ is a subset of $U$ and a sampling design $p(.)$ is a probability distribution on all the subsets of $U$. A random sample $S$ is a vector that maps all subsets $s$ to an $N$ vector of 0 or 1 such that
	$$
	P(S = s) = p(s),~ p(s)\ge 0, \mbox{ and } \sum_{s\subset U}p(s) = 1, \mbox{ for all } s\subset U.
	$$
	
	In order to estimate this parameter, two approaches can be used
	\begin{enumerate}
		\item Consider a sampling design $p(.)$ and find a sampling method (or algorithm) to implement this design. Based on this approach, according to $p(.)$, all the first and second-order inclusion probabilities can be specified using
		$$
		\begin{array}{lll}
		\pi_k=\textrm{Pr}(k\in S)=\sum_{s\ni k} p(s) \mbox{ and }\\
		\pi_{k\ell}=\textrm{Pr}(k,\ell\in S)=\sum_{s\ni\{k,\ell\} } p(s), \mbox{ for all } k,\ell \in U.
		\end{array}
		$$
		
		\item Consider a vector of first-order inclusion probabilities $\bm{\pi}$ and find a sampling method to select a random sample $S$ such that $\textrm{Pr}(k\in S)=\pi_k,$ for all $k\in U$.
	\end{enumerate}
	
	In this article, the second approach is considered. If $\pi_k>0$, for all $k\in U$, $Y$ can be estimated unbiasedly by means of the first-order inclusion probabilities using the expansion estimator~\citep{nar:51,horvitz1952generalization}
	\begin{equation*}
		\widehat{Y}=\sum_{k\in S}\frac{y_k}{\pi_k}.
	\end{equation*}
	Its precision depends on the second-order inclusion probabilities
	\begin{equation*}
		\textrm{Var}(\widehat{Y})=\sum_{k\in U}\sum_{\ell\in U} (\pi_{k\ell}-\pi_k\pi_\ell)\frac{y_k}{\pi_k}\frac{y_\ell}{\pi_\ell}.
	\end{equation*}
	
	Indeed, the values of $\bm{\pi}$ indicates the desire chance (probability) for the units to be selected in the final sample. These probabilities can be calculated based on some proper auxiliary variables. If $\pi_{k\ell}>0,$ for all $k,\ell\in U,$  $\textrm{Var}(\widehat{Y})$ can be unbiasedly estimated by
	\begin{equation}
		\widehat{\textrm{Var}}(\widehat{Y})=\sum_{k\in S}\sum_{\ell\in S} \frac{\pi_{k\ell}-\pi_k\pi_\ell}{\pi_{k\ell}} \frac{y_k}{\pi_k}\frac{y_\ell}{\pi_\ell}\nonumber,
	\end{equation}
	however this does not guaranty to have a precise estimation. According to the sampling method, the precision of the expansion estimator can be affected. As the last point in this section, it is notable that precision is not always the most important criteria to evaluate a strategy (set of a sampling method and an estimator). Other important criteria can be applicability of the strategy according to accessibility of the data over time that will be discussed more in Section~\ref{streams}.

	
	\section{One-Step One-Decision Sampling Method} \label{MainD}
	
	In this section, we present the procedure called One-Step One-Decision (OSOD) sampling method. The general idea of the method is that at each step a complete decision is made on the considered unit. This means that the unit is selected in the sample or not. After that, the inclusion probabilities are updated according to the decision made on this unit. This procedure is then repeated on the following units and it takes at most $N$ steps to obtain the final sample.
	
	Consider a finite population $U$. Define the sum of the inclusion probabilities
	$$
	n=\sum_{k\in U}\pi_k,
	$$
	that can be non-integer. For simplicity, we first consider the method where the decision is made for the first unit in the list. Suppose also that
	\begin{equation*}
		\sum_{k=1}^N\pi_k\ge 1 \mbox{ and } \sum_{k=1}^N(1-\pi_k)\ge 1.
	\end{equation*}
	The first unit is selected with probability $\pi_1$. Let $\pi_1^1$ be the updated inclusion probability of the unit 1. It can be 0 or 1 and can be summarize as follow
	\begin{eqnarray}\nonumber
		\pi^1_1=\left\{\begin{array}{ll}
			1& \mbox{ with probability } \pi_1 \\
			0 & \mbox{ with probability } 1-\pi_1.
		\end{array}\right.
	\end{eqnarray}
	After deciding about the first unit, the remaining inclusion probabilities need to be updated. The whole idea of the method is to increase (respectively decrease) inclusion probabilities of the following units depending if $\pi_1^1$ has been changed to 0 (respectively 1). The inclusion probabilities of the remaining units are updated as follow
	\begin{equation*}
		\pi^1_k=\left\{
		\begin{array}{ll}
			\pi^1_k(0)= \min(c_1\pi_k,1) & \mbox{ if } \pi^1_1=0\\[2mm]
			\displaystyle\pi^1_k(1)=\frac{\pi_k-\pi^1_k(0)(1-\pi_1)}{\pi_1} & \mbox{ if } \pi^1_1=1,
		\end{array}
		\right.
		\text{ for } k = 2,\dots,N,
	\end{equation*}
	where constant $c_1$ is defined by
	\begin{equation}\label{eq:prop}
		\sum_{k=2}^N \min(c_1\pi_k,1)=n.
	\end{equation}
	Equation \eqref{eq:prop} shows that the inclusion probabilities of the remaining units are increased proportionally such that the sample size is still respected. After the first step, the decision is irrevocably made for unit 1. Simply, this operation can be repeated for other steps $t=2,\dots,N.$ At step $t$, decisions are made about the first $t-1$ units and $\bm{\pi}$ has been updated $t-1$ times. The updated vector is denoted by
	$$
	\bm{\pi}^{t-1}=(\pi^{t-1}_1,\pi^{t-1}_2,\dots,\pi^{t-1}_k,\dots,\pi^{t-1}_N)^\top,
	$$
	where its first $t-1$ units are in $\{0,1\}$.
	Again, it is supposed that
	\begin{equation}\label{conditions}
		\sum_{k=t}^{N}\pi^{t-1}_k\ge 1 \mbox{ and }\sum_{k=t}^{N}(1-\pi^{t-1}_k)\ge 1,
	\end{equation}
	and a decision is taken for the $t$th unit with probability
	\begin{eqnarray}\nonumber
		\pi^{t}_t=\left\{
		\begin{array}{ll}
			1& \mbox{ with probability } \pi^{t-1}_t\\
			0 & \mbox{ with probability }1-\pi^{t-1}_t.
		\end{array}
		\right.
	\end{eqnarray}
	Then, the inclusion probabilities are updated by
	\begin{equation*}
		\pi^t_k=\left\{
		\begin{array}{ll}
			\pi^t_k(0)=\min(c_t\pi^{t-1}_k,1)& \mbox{ if } \pi^{t}_t=0 \\[2mm]
			\displaystyle \pi^t_k(1)=\frac{\pi^{t-1}_k-\pi^{t}_k(0)(1-\pi^{t-1}_t)}{\pi^{t-1}_t} & \mbox{ if } \pi^t_t=1,
		\end{array}\right.
		\text{ for } k=t+1,t+2\dots,N,
	\end{equation*}
	where $c_t$ is defined by
	$$\sum_{k=t+1}^N \min(c_t\pi^{t-1}_k,1)=n-n_t,$$
	and $n_t$ is the number of selected units up to step $t$.

Conditions (\ref{conditions}) are necessary to leave the possibility of updating the other inclusion probabilities. For example at the first step it is required to have enough space in other inclusion probabilities to distribute $\pi_1$  as $\pi^1_k(0)=\pi_k+\alpha_k\pi_1$ (if unit $1$ is not selected) or to remove $1-\pi_1$ from them as $\pi^1_k(1)=\pi_k-\alpha_k(1-\pi_1)$ (if unit $1$ is selected) for some $0<\alpha_k\le1$ on the following inclusion probabilities. Then, hereafter, OSOD stands for the design with a population that satisfies Conditions (\ref{conditions}).
	
Result~\ref{res3} shows that the sampling procedure respects the inclusion proba\-bilities and thus the fixed sample size. The procedure is presented more formally in Algorithm~\ref{alg1}. Currently, the sampling procedure takes into account the whole population to modify the inclusion probabilities. In the next section, a great improvement for sampling from a stream is proposed. Indeed, only a small part of the following units can be considered to update the inclusion probabilities. This completely modifies the application of the method.
	
	\begin{res}\label{res3}
		With OSOD, for all $t = 1,2,\dots$ we have
		$$
		\textrm{E}(\pi^{t}_k)=\pi_k,\;\mbox{ for all }k\in U,\;\;\;\;\;\;\mbox{and}\;\;\;\;\;\; \sum_{k=1}^N\pi^t_k=\sum_{k=1}^N\pi_k.$$
	\end{res}
	\noindent The proof is given in Appendix.

	%

	\begin{algorithm}
		\caption{One-Step One-Decision sampling algorithm with parameter $\bm{\pi}$.}\label{alg1}
		\begin{algorithmic}
			\footnotesize
			\State Initialize with $n_0=0$, $\bm{\pi}^0=\bm{\pi}$ and $n=\sum_{k\in U}\pi_k$.
			\For{$t=1,2,3,\dots,N$}
			\State Calculate $c_t$ such that $\displaystyle\sum_{k=t+1}^N\min(c_t\pi^{t-1}_k,1)=n-n_t$ where $\displaystyle n_t=\sum_{k=1}^{t-1}\pi^{t-1}_k$,
			\State Generate $u$, a realization of a uniform random variable in $[0,1]$,
			\If{$u> \pi^{t}_t$}
			\State Set $\pi^t_t=0$ and $\pi^{t}_k=\min(c_t\pi^{t-1}_k,1)$ for all $k>t$.
			\Else
			\State Set $\pi^t_t=1$ and $\pi^{t}_k=\frac{\pi^{t-1}_k-\min(c_t\pi^{t-1}_k,1)(1-\pi^{t-1}_t)}{\pi^{t-1}_t}$ for all $k>t$.
			\EndIf
			\If{ $\pi^{t}_k\in \{0,1\}$ for all $k=t+1,t+2,...,N$}
			\State Stop
			\EndIf
			\EndFor
			%
			%
			%
			%
		\end{algorithmic}
	\end{algorithm}

	
	\section{Extending the Method to Stream Sampling}\label{streams}

	A data stream is a sequence of information arriving sequentially in time. When sampling streams, usually a very large set of data coming in continuously. Because of the limited space needed to store the data, it is desirable to decide on the selection of units at the right time.
	
	As pointed in Section~\ref{intro}, in the methods proposed so far, the decision about a unit that is at the beginning of a stream can be made very late. Ideally, the decision to select a unit or not from a stream must be taken as soon as possible, which the OSOD method allows. The idea to extend the OSOD method to stream sampling is to consider a window of units instead of the entire population to update inclusion probabilities. Result \ref{res1} shows that the number of units to consider depends on the inclusion probabilities on the following units. Essentially, it is enough to wait until having a set of units such that there exists a constant $c_t$ that satisfies $\pi_t \geq (1-1/c_t)$. Moreover, Result \ref{resMain} shows also that a sufficient condition to update inclusion probabilities is that the inclusion probabilities within the window sum up to integer.
	
	\begin{res}\label{res1}
		In OSOD, a necessary and sufficient condition for all $\pi^t_k(1)$ to be non-negative is that
		\begin{equation*}\label{maincondition}
			\pi_t\geq 1-1/c_t.
		\end{equation*}
	\end{res}
	
	\begin{res}\label{resMain}
		In OSOD, a sufficient condition for all $\pi^t_k(1)$ to be non-negative is that $n=\sum_{k\in U}\pi_k$ is integer.
	\end{res}
	\noindent The proofs are given in Appendix.

	Hence after making decision about the first unit, enough units with a $c_1$ that satisfies $\pi_1\geq(1-1/c_1)$ is required. Also if there is an integer window then it is possible to apply the method before. Still, it could be possible that there is no window that sums up to integer and no constant $c_1$ that satisfies the condition until the end of the stream. In this case, it is possible to use Result \ref{resMain} and the notion of phantom unit~\citep{graf:mat:qua:til:12} to artificially transform the sum of the inclusion probabilities to an integer and finish the procedure. This step is well explained in the Section \ref{sec:nonint}. The method has many advantages for performing sampling in streams, including
	\begin{itemize}
		\item The decision about the units can be made based on their order,
		\item To update the population inclusion probabilities after selecting a unit, the whole population is not needed and only a small part of it (based on Result~\ref{res1} or Result~\ref{resMain}) is enough. In fact, in the case of streams, there is usually no finite set called population and the data is generated online,
		\item It is possible to make decision about one part of the population indepen-dently of the other parts.
	\end{itemize}
	Algorithm \ref{alg2} gives a detailed explanation of the proposed method. The next two Sections discuss example and show intuitively why the proposed method spread the units according to their order and the case where no constant $c_t$ can be found for the entire population.
	
	%

	\begin{algorithm}
		\caption{One-Step One-Decision Stream Sampling algorithm.}\label{alg2}
		\begin{algorithmic}
			\State Initialize with $\pi_1^0 = \pi_1$,
			\footnotesize
			\For{$t=1,2,3,\dots$}
			\State Consider unit $t$,
			\If {There is a window of units of size $m$ for which $\pi^{t-1}_t\ge (1-1/c_t)$}
			\State 1) Set $n-n_t=\displaystyle\sum_{k=t}^{t+m-1}\pi^{t-1}_k$, $N=t+m-1$,
			\State 2) Implement \textbf{Algorithm~\ref{alg1}}, just for one step at $t$.
			\Else
			\State 1) Consider the largest possible window (all the available data), say of size $M$,
			\State2) Add a phantom unit $k=t+M$ with $$\pi^{t-1}_{t+M}=1-\left(\sum_{k=t}^{t+M-1 }\pi^{t-1}_k-\left\lfloor\sum_{k=t}^{t+M-1 }\pi^{t-1}_k\right\rfloor\right),$$
			\State 3) Set $n_t=0$, $n=\displaystyle\sum_{k=t}^{t+M }\pi^{t-1}_k$ and $N=t+M$,
			\State 4) Implement \textbf{Algorithm~\ref{alg1}} from step $t$ for $M$ steps,
			\State 5) Delete the phantom unit with $\pi^{t+M-1}_{t+M}$,
			\State 6) Set $t=t+M$.
			
			\EndIf
			\EndFor
		\end{algorithmic}
	\end{algorithm}

	
	\section{Spread the Units According to their Order}\label{sec:spread}
	
	For some reasons, such as a high correlation between the main variable and the indices, it is often desirable to select the sample spread on the order of the indices. Let $m$ be the number of units in the window satisfying the condition of Result~\ref{res1}. Except for very extreme inclusion probabilities, it appears easy to find small $m$ in Algorithm~\ref{alg2}.
	
	
	To spread the sample, the parameter $m$ in Algorithm~\ref{alg2} is set as small as possible. Select the smallest window in the sampling procedure is somehow evocative of local pivotal method~\citep{gra:lun:sch:12}. Indeed with such strategy, the respective unit fights with some closest units around and has its inclusion probability modified to 1 or 0. This idea of creating repulsion between nearest units is exactly the same if a small $m$ is chosen. Indeed, for a particular unit, if the inclusion probability is changed to 1, the procedure will decrease inclusion probabilities of the following neighbours. If the window is small, the next unit is more likely to be modified directly to 0 or in the next step. Then $m$ can be considered as a leverage to control the level of spread of the sample. Again, one of the advantages of OSOD over the classical methods is that it does not postpone the decision on the units.
	
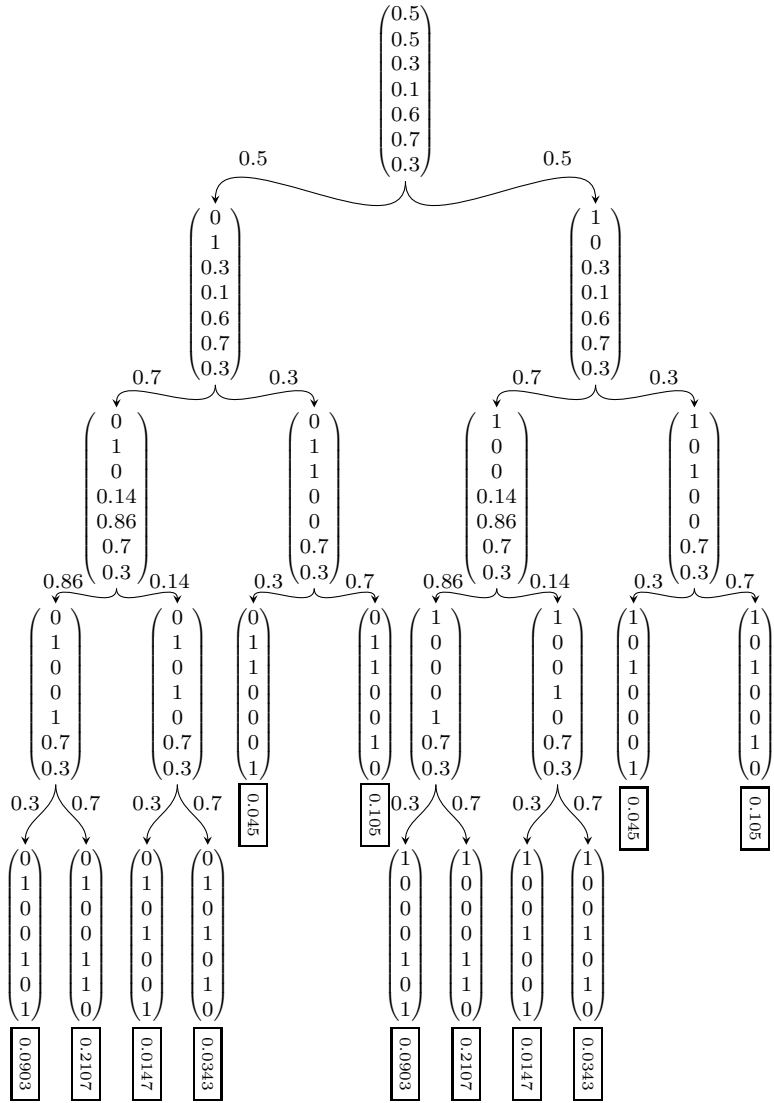
\begin{figure}[htb!]
		\centering
		\begin{tikzpicture}[scale=1]
			\node at (10,10.2) {\scriptsize\vva};
			\node at (7.5,7.5) {\scriptsize\vvb};\node at (12.5,7.5) {\scriptsize\vvc};
			\node at (6.2,4.8) {\scriptsize\vvd};\node at (8.8,4.8) {\scriptsize\vve};\node at (11.2,4.8) {\scriptsize\vvf};\node at (13.8,4.8) {\scriptsize\vvg};
			\node at (5.4,2.2) {\scriptsize\vvh};\node at (7.0,2.2) {\scriptsize\vvi};\node at (8.0,2.2) {\scriptsize\vvj};\node at (9.6,2.2) {\scriptsize\vvk};
			\node at (10.4,2.2) {\scriptsize\vvl};\node at (12,2.2) {\scriptsize\vvm};\node at (13.0,2.2) {\scriptsize\vvn};\node at (14.6,2.2) {\scriptsize\vvo};
			\node at (5.0,-1) {\scriptsize\vvp};\node at (5.8,-1) {\scriptsize\vvq};
			
			\node at (6.6,-1) {\scriptsize\vvr};\node at (7.4,-1) {\scriptsize\vvs};
			
			\node at (10.0,-1) {\scriptsize\vvt};\node at (10.8,-1) {\scriptsize\vvu};
			
			\node at (11.6,-1) {\scriptsize\vvv};\node at (12.4,-1) {\scriptsize\vvw};
			\draw[->,>=stealth,thin] (10,9.0) to [out=270,in = 90] (7.5,8.7);
			\draw[->,>=stealth,thin] (10,9.0) to [out=270,in = 90] (12.5,8.7);
			\draw[->,>=stealth,thin] (7.5,6.3) to [out=270,in = 90] (6.2,6); 
			\draw[->,>=stealth,thin] (7.5,6.3) to [out=270,in = 90] (8.8,6); 
			\draw[->,>=stealth,thin] (12.5,6.3) to [out=270,in = 90] (11.2,6);
			\draw[->,>=stealth,thin] (12.5,6.3) to [out=270,in = 90] (13.8,6);
			\draw[->,>=stealth,thin] (6.2,3.6) to [out=270,in = 90] (5.4,3.4); 
			\draw[->,>=stealth,thin] (6.2,3.6) to [out=270,in = 90] (7.0,3.4); 
			\draw[->,>=stealth,thin] (8.8,3.6) to [out=270,in = 90] (8.0,3.4);	
			\draw[->,>=stealth,thin] (8.8,3.6) to [out=270,in = 90] (9.6,3.4);	
			\draw[->,>=stealth,thin] (11.2,3.6) to [out=270,in = 90] (10.4,3.4);	
			\draw[->,>=stealth,thin] (11.2,3.6) to [out=270,in = 90] (12,3.4);	
			\draw[->,>=stealth,thin] (13.8,3.6) to [out=270,in = 90] (13,3.4);	
			\draw[->,>=stealth,thin] (13.8,3.6) to [out=270,in = 90] (14.6,3.4);
			\draw[->,>=stealth,thin] (5.4,1.0) to [out=270,in = 90] (5.0,0.2); 
			\draw[->,>=stealth,thin] (5.4,1.0) to [out=270,in = 90] (5.8,0.2); 
			\draw[->,>=stealth,thin] (7.0,1.0) to [out=270,in = 90] (6.6,0.2); 
			\draw[->,>=stealth,thin] (7.0,1.0) to [out=270,in = 90] (7.4,0.2); 
			\draw[->,>=stealth,thin] (10.4,1.0) to [out=270,in = 90] (10.0,0.2);	
			\draw[->,>=stealth,thin] (10.4,1.0) to [out=270,in = 90] (10.8,0.2);	
			\draw[->,>=stealth,thin] (12,1.0) to [out=270,in = 90] (11.6,0.2);	
			\draw[->,>=stealth,thin] (12,1.0) to [out=270,in = 90] (12.4,0.2);	
			
			\node at (8,9.3) {\scriptsize 0.5};\node at (12,9.3) {\scriptsize 0.5};
			\node at (6.6,6.4) {\scriptsize 0.7};\node at (8.4,6.4) {\scriptsize 0.3};\node at (11.6,6.4) {\scriptsize 0.7};\node at (13.4,6.4) {\scriptsize 0.3};
			
			\node at (5.5,3.7) {\scriptsize 0.86};\node at (6.9,3.7) {\scriptsize 0.14};
			\node at (8.2,3.7) {\scriptsize 0.3};\node at (9.4,3.7) {\scriptsize 0.7};
			\node at (10.5,3.7) {\scriptsize 0.86};\node at (11.9,3.7) {\scriptsize 0.14};
			\node at (13.2,3.7) {\scriptsize 0.3};\node at (14.4,3.7) {\scriptsize 0.7};
			
			\node at (5.0,.75) {\scriptsize 0.3};\node at (5.8,.75) {\scriptsize 0.7};
			\node at (6.6,.75) {\scriptsize 0.3};\node at (7.4,.75) {\scriptsize 0.7};

			\node at (10,0.75) {\scriptsize 0.3};
			\node at (10.8,.75) {\scriptsize 0.7};
			\node at (11.6,.75) {\scriptsize 0.3};
			\node at (12.4,.75) {\scriptsize 0.7};
			
			\node[rotate = -90] at (8.0,0.6) {\tiny\boxed{{0.045}} };
			\node[rotate = -90] at (9.6,0.6) {\tiny\boxed{{0.105}}};
			\node[rotate = -90] at (13,0.55) {\tiny\boxed{{0.045}}};
			\node[rotate = -90] at (14.6,0.55) {\tiny \boxed{{0.105}}};
			
			\node[rotate = -90] at (5.0,-2.7) {\tiny \boxed{{0.0903}}};	
			\node[rotate = -90] at (5.8,-2.7) {\tiny \boxed{{0.2107}}};		
			\node[rotate = -90] at (6.6,-2.7) {\tiny \boxed{{0.0147}}};	
			\node[rotate = -90] at (7.4,-2.7) {\tiny \boxed{{0.0343}}};		
			\node[rotate = -90] at (10.0,-2.7) {\tiny \boxed{{0.0903}}};	
			\node[rotate = -90] at (10.8,-2.7) {\tiny \boxed{{0.2107}}};		
			\node[rotate = -90] at (11.6,-2.7) {\tiny \boxed{{0.0147}}};	
			\node[rotate = -90] at (12.4,-2.7) {\tiny \boxed{{0.0343}}};	
		\end{tikzpicture}
		\caption{Well spread strategy of OSOD on  $\pi=(0.5,0.5,0.3,0.1,0.6,0.7,0.3)^\top$ with considering three small windows ($m_1=2, m_2=3$ and $m_3=2$) on the population in Equation~(\ref{wellspread}). The boxed numbers indicate the design of all the possible samples. Each time at least a complete decision about one unit is taken and then just the inclusion probabilities inside the respective window are updated (and not the whole population). \label{figWS} }
	\end{figure}

	
	To clarify this strategy, consider $$\bm{\pi}=(0.5,0.5,0.3,0.1,0.6,0.7,0.3)^\top.$$
	Based on Result~\ref{resMain}, there are at least three windows to make decision about the first unit, $m=2,5,7$ (as the sum is integer for each of them). More formally the problem can be presented as
	\begin{equation}\label{wellspread}
		\bm{\pi}=(
			\underbrace{\underbrace{\underbrace{ 0.5,0.5,}_{\bm{w}_1,\text{ with $m=2$}}\;\;\;\;\;\;
		\underbrace{ 0.3,0.1,0.6,}_{\bm{w}_2,\text{ with $m=3$}}}_{\bm{w}_1,\text{ with $m=5$}}\;\;\;\;\;\;
		\underbrace{ 0.7,0.3}_{\bm{w}_3,\text{ with $m=2$}}
	}_{\bm{w}_1,\text{ with $m=7$}})^\top.
	\end{equation}
	If $m=2$, then one of the first two units will be selected in the sample. But if $m=7$ is considered, it is possible to have both, or none of the first units inside the final sample. Figure~\ref{figWS} shows the sampling process of the method with the small windows ($m=2,\; m=3$ and $m=2$) strategy. For all the cases, the sample is well spread in the three windows.
	
	Considering all the population as a window (i.e. $m = 7$) and apply the method, will lead to obtain one of the sample in Table~\ref{tab1} (probabilities are based on 10,000 iterations). In this situation, it is possible to have a sample that is not spread over the population at all, such as samples that are labelled in the table by indices 6, 16, and 31.

	\begin{table}[htb!]
		\begin{center}
			\caption{Implementing OSOD on $\bm{\pi}=(0.5,0.5,0.3,0.1,0.6,0.7,0.3)^\top$ with considering the whole population as the only window. The design is calculated based on 10,000 iterations.}\label{tab1}
		\end{center}
		\begin{center}
			\small
			\begin{tabular}{ccccccccc}
				\hline
				No. of Sample	&	\multicolumn{7}{c}{The Samples}	&	Sampling Design	\\ \hline
				1	&	0	&	0	&	1	&	0	&	0	&	1	&	1	&	0.0123	\\
				2	&	0	&	0	&	1	&	0	&	1	&	0	&	1	&	0.0069	\\
				3	&	0	&	0	&	1	&	0	&	1	&	1	&	0	&	0.0369	\\
				4	&	0	&	0	&	1	&	1	&	0	&	0	&	1	&	0.0078	\\
				5	&	0	&	0	&	1	&	1	&	0	&	1	&	0	&	0.0148	\\
				6	&	0	&	0	&	1	&	1	&	1	&	0	&	0	&	0.0144	\\
				7	&	0	&	1	&	0	&	0	&	0	&	1	&	1	&	0.0582	\\
				8	&	0	&	1	&	0	&	0	&	1	&	0	&	1	&	0.0308	\\
				9	&	0	&	1	&	0	&	0	&	1	&	1	&	0	&	0.1537	\\
				10	&	0	&	1	&	0	&	1	&	0	&	0	&	1	&	0.0024	\\
				11	&	0	&	1	&	0	&	1	&	0	&	1	&	0	&	0.0039	\\
				12	&	0	&	1	&	0	&	1	&	1	&	0	&	0	&	0.0032	\\
				13	&	0	&	1	&	1	&	0	&	0	&	0	&	1	&	0.0266	\\
				14	&	0	&	1	&	1	&	0	&	0	&	1	&	0	&	0.0651	\\
				15	&	0	&	1	&	1	&	0	&	1	&	0	&	0	&	0.0554	\\
				16	&	0	&	1	&	1	&	1	&	0	&	0	&	0	&	0.0027	\\
				17	&	1	&	0	&	0	&	0	&	0	&	1	&	1	&	0.0804	\\
				18	&	1	&	0	&	0	&	0	&	1	&	0	&	1	&	0.0415	\\
				19	&	1	&	0	&	0	&	0	&	1	&	1	&	0	&	0.1953	\\
				20	&	1	&	0	&	0	&	1	&	0	&	0	&	1	&	0.0074	\\
				21	&	1	&	0	&	0	&	1	&	0	&	1	&	0	&	0.0171	\\
				22	&	1	&	0	&	0	&	1	&	1	&	0	&	0	&	0.0178	\\
				23	&	1	&	0	&	1	&	0	&	0	&	0	&	1	&	0.0085	\\
				24	&	1	&	0	&	1	&	0	&	0	&	1	&	0	&	0.0216	\\
				25	&	1	&	0	&	1	&	0	&	1	&	0	&	0	&	0.0186	\\
				26	&	1	&	0	&	1	&	1	&	0	&	0	&	0	&	0.0030	\\
				27	&	1	&	1	&	0	&	0	&	0	&	0	&	1	&	0.0153	\\
				28	&	1	&	1	&	0	&	0	&	0	&	1	&	0	&	0.0351	\\
				29	&	1	&	1	&	0	&	0	&	1	&	0	&	0	&	0.0324	\\
				30	&	1	&	1	&	0	&	1	&	0	&	0	&	0	&	0.0048	\\
				31	&	1	&	1	&	1	&	0	&	0	&	0	&	0	&	0.0061	\\
				\hline
			\end{tabular}%
		\end{center}
	\end{table}

	
	\section{Populations with Non-Integer Sum of Inclusion Probabilities}\label{sec:nonint}
	
	As pointed before, the method is perfectly working if the sum of inclusion probabilities within the window is not an integer as long as the condition of Result \ref{res1} is satisfied. An issue occurs if the condition is not satisfied and the sum of the inclusion probabilities is not an integer. In such situation, this problem can be solved by using the notion of phantom unit~\citep{graf:mat:qua:til:12}. The ingenuity is to add a phantom unit that has inclusion probability equal to the difference needed to reach the ceiling integer. Then, by Result \ref{resMain}, the method is working and at the end of the process the phantom unit will be removed of the sample. The strategy respects the inclusion probabilities. As an example, consider the following inclusion probability vector
	%
	
	\begin{eqnarray}\nonumber
		\small
		\begin{array}{ll}
			\bm{\pi}=(0.85,0.90,0.90,0.02,00.02,0.98,0.99,0.95,0.99,0.01,0.01,0.99,0.99,0.99)^\top.
		\end{array}
	\end{eqnarray}
	In this case, because of extreme inclusion probabilities, the condition of Result \ref{res1} is not satisfied on the whole population and the sum of the inclusion probabilities is not an integer. With implementing the method, the updated inclusion probabi\-lities are (rounded to 2 decimal places)

	\begin{eqnarray}\nonumber
		\bm{\pi}^1=\small\left\{
		\begin{array}{llll}
			(0.00,\;1.00,\; 1.00,\;  0.20,\;  0.20,\; 1.00,\; 1.00,\;\\\hspace{1cm} 1.00,\; 1.00,\;  0.10,\;  0.10,\;  1.00,\;  1.00,\;  1.00)^\top& \mbox{ if } \pi^{1}_1=0 \\\\
			\displaystyle (1.00, 0.88, 0.88, -0.01, -0.01, 0.98, 0.99,\\\hspace{1cm} 0.94, 0.99, -0.01, -0.01,  0.99,  0.99,  0.99)^\top & \mbox{ if } \pi^1_1=1,
		\end{array}\right.
	\end{eqnarray}
	Because of negative values, $\bm{\pi}^1(1)$ is not acceptable. As the sum of inclusion probabilities is $9.59$, the proposed solution is to add a phantom unit with inclusion probability $\pi^*=0.41$. Based on Result~\ref{resMain} it is possible to apply the method on
	
	which leads to updated vectors as
	
	\begin{eqnarray}\nonumber
		\bm{\pi}^{*1}=\small\left\{
		\begin{array}{llll}
			(0.00, 1.00, 1.00, 0.04, 0.04, 1.00, 1.00,\\\hspace{1cm} 1.00, 1.00, 0.02,  0.02,  1.00,  1.00,  1.00,  \boxed{0.87})^\top& \mbox{ if } \pi^{1}_1=0 \\\\
			\displaystyle (1.00, 0.88, 0.88, 0.02, 0.02, 0.98, 0.99,\\\hspace{1cm} 0.94, 0.99, 0.01,  0.01,  0.99,  0.99,  0.99,  \boxed{0.33})^\top & \mbox{ if } \pi^1_1=1.
		\end{array}\right.
	\end{eqnarray}
	
	As the sum remains constant in the updated inclusion probabilities, the method can be applied step by step to decide about all the units. At the last step, the phantom unit will be ignored. With this procedure, inclusion probabilities and also the expectation of the sample size are respected. With this strategy it is always possible to finish the selection of the sample even if the two necessary conditions are not fulfilled. It is important to note that if this strategy is applied while the selection process is not at the end of its course, the sample size is random with expectation equal to the sum of the inclusion probabilities. For more information, the reader can find details in~\citep{graf:mat:qua:til:12}.

	\section{Simulations}\label{sec:simu}
	
\begin{figure}[ht!]
		\centering
		\input{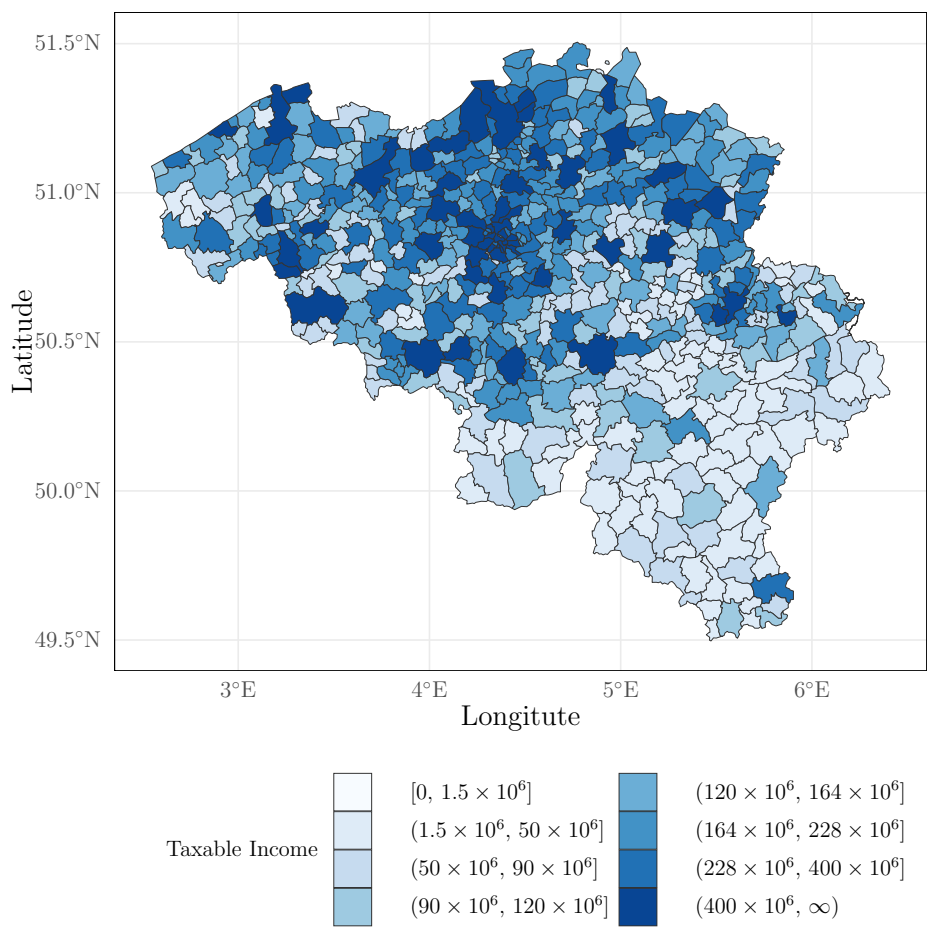}
		\caption{Belgian municipalities dataset. The discrete gradient scale represents the taxable income in euros. }
		\label{fig:belg}
	\end{figure}

 	As the method has significant new advantages over other methods, we show on the basis of a real dataset, that in terms of estimation accuracy, the method is comparable to its competitors. 
 	The data set is available in the \texttt{R} package \texttt{sampling}~\citep{TilMat2015} under the name \texttt{belgianmunicipalities}. It contains different information about the Belgian population between 2001 and 2004 by municipalities. The data set contains 589 rows corresponding to each municipality of Belgium. Here just two variables were considered, the variable of interest $y$ which represents the taxable income in euros in 2001, and the auxiliary variable $x$ which provides the total population on July 1, 2004. Figure \ref{fig:belg} shows the variable of interest by municipalities.
	
	Next, based on 10,000 simulations, several sampling methods were considered and expansion estimator for each of them was calculated. OSOD method was compared to five other sampling methods, random systematic~\citep{mad:49}, random pivotal~\citep{dev:til:98}, Till\'e's method~\citep{til:96a}, Midzuno's method~\citep{mid:50} and finally the conditional Poisson sampling also called the maximum entropy method~\citep{che:liu:97}. The reader can find more details on each of these methods in~\citep{til:06}.
	
\begin{figure}[ht!]
		\centering
		\input{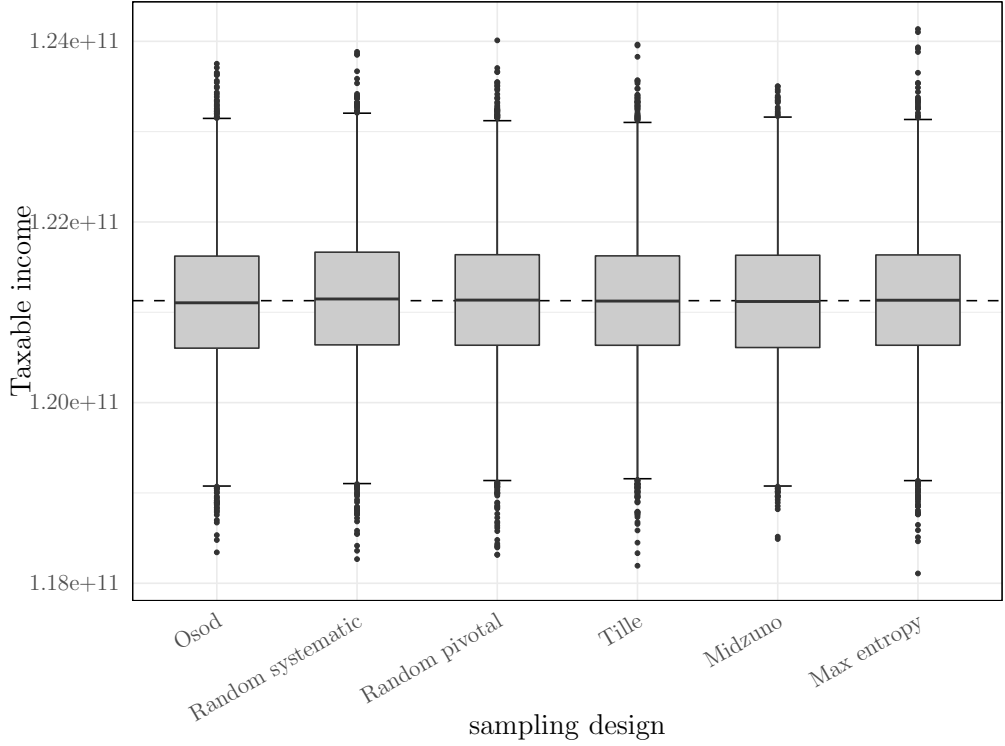}
		\caption{Boxplot of the Horvitz-Thompson estimator for the variable of interest taxable income of the Belgian municipalities dataset. Different sampling design are illustrated as well as the proposed method. We find that the proposed sampling design is quite comparable in terms of variance estimator.}
		\label{fig:boxplot}
	\end{figure}

	The sample size was set to 200 municipalities and the inclusion probabilities were unequal and set proportional to the total population variable $x$. Figure \ref{fig:boxplot} shows the boxplot of the expansion estimator of the total for each sampling design and confirms that our method is completely comparable in terms of variance and respect the unequal inclusion probabilities. The method was imple-mented in the \texttt{R} package \texttt{StratifiedSampling}~\citep{StratifiedSampling} under the name \texttt{osod}.



	\section{Conclusion}

	The OSOD method is the first method in unequal probability sampling that can decide about the units completely based on their order. The method is flexible on the order of the data and spreading the data over the population. The order can be arbitrary but if the data are the result of a stream, the method can be very useful to decide about what units should be stored as a sample over time. Then after deciding about each unit to update the inclusion probabilities, there is no need to use all the population and usually just a small window of the population can be enough. Also based on the size of the windows, it is possible to control sample rate of spreading over the population indices. The indices can indicate time or location of the data. If the main variable is correlated with the indices, it would be very beneficial to select the windows as small as possible.
	
	\section*{Appendix}
	
	\begin{proof}[\textbf{\bf Proof of Result~\ref{res3}}]
		Without loss of generality and for ease of notation we only do the proof for $t = 1$. For $k=1$ it is obvious that $\textrm{E}(\pi^1_1)=\pi_1$, and for $k=2,3,\dots,N$, 
		\begin{eqnarray}
			\textrm{E}(\pi^1_k) =\pi^1_k(0)(1-\pi_1)+\pi^1_k(1)\pi_1=\pi_k^1(0)(1-\pi_1)+\frac{\pi_k-\pi^1_k(0)(1-\pi_1)}{\pi_1}\pi_1=\pi_k.\nonumber
		\end{eqnarray}
		Also for the sum of the inclusion probabilities, we have
		$$
		\sum_{k\in U}\pi_k^1(0) = 0+\sum_{k=2}^{N}\pi^1_k(0)=\sum_{k=2}^{N}\min(c_1\pi_k,1)=n=\sum_{k=1}^{N}\pi_k,
		$$
		and
		$$\begin{array}{lll}
			\displaystyle\sum_{k\in U}\pi_k^1(1) &=&\displaystyle 1+\sum_{k=2}^{N}\pi^1_k(1)\\
			&=&\displaystyle1+\frac{\sum_{k=2}^{N}\pi_k-(1-\pi_1)\sum_{k=2}^{N}\pi^1_k(0)}{\pi_1}\\
			&=&\displaystyle\frac{\pi_1+\{(n-\pi_1)-n+n\pi_1\}}{\pi_1}\\
			&=&n\\
			&=&\displaystyle\sum_{k=1}^{N}\pi_k.
		\end{array}$$
		
	\end{proof}

	\begin{proof}[\textbf{\bf Proof of Result~\ref{res1}}]
		Without loss of generality and for ease of notation we only do the proof for $t = 1$. Since 
		\begin{eqnarray}\nonumber
			\pi^1_k(1)=\left\{
			\begin{array}{ll}
				\displaystyle\frac{\pi_k-\min(c_1\pi_k,1)(1-\pi_1)}{\pi_1}\le \frac{\pi_k-c_1\pi_k(1-\pi_1)}{\pi_1} & \mbox{ if }\pi^1_k(1)=1\\[3mm]
				\displaystyle\frac{\pi_k-c_1\pi_k(1-\pi_1)}{\pi_1} &\mbox{ if } \pi_k^1(1)<1,
			\end{array}\right.
		\end{eqnarray}
		for $k=2,3,\dots,N$,
		a necessary and sufficient condition is that $\pi_1\geq 1-1/c_1$, which gives the result.
	\end{proof}

	\begin{proof}[\textbf{\bf Proof of Result~\ref{resMain}}]
		Let $$U_{A} = \{k\in U|0< \pi_k^1(0)<1\},\; U_{B} = \{k\in U|\pi_k^1(0)=1\},$$
		$$A=\sum_{k\in U_A}\pi_k
		\mbox{ and }B=\sum_{k\in U_{B}}\pi_k.$$
		Then it is possible to decompose $n$ as
		\begin{equation}\label{Yp1}
			\pi_1+A+B=n
		\end{equation}
		and
		$$c_1A+\#U_B=n.$$
		Now, if $n$ is integer, $c_1A$ is an integer denoted by $d$. Thus $A=d/c_1$. In this case, $\#U_B=n-d$. Moreover, it is easy to see that $B\le \#U_B$. Now, from Equation~\eqref{Yp1}, we have
		$$\pi_1=n-A-B=n-\frac{d}{c_1}-B\ge n-\frac{d}{c_1}-\#U_B=n-\frac{d}{c_1}-(n-d)=d\left(1-\frac{1}{c_1}\right).$$
		Now if $d=1,2,\dots$ then $\pi_1\ge (1-1/c_1)$ and if $d=0$ then $\#U_B=n$ or in other words, $\pi^1_k(0)=1$ for all $k=2,3,\dots,N$. Therefore to have all $\pi^1_k(1)\ge 0$ we need to have
		\begin{equation}\label{comproof}
			\pi^1_k(1)=\frac{\pi_k-(1-\pi_1)}{\pi_1}\ge 0, \Rightarrow \pi_k+\pi_1-1\ge 0.
		\end{equation}
		But as in such cases, $\pi^1_k(0)=\pi_k+\alpha_k \pi_1=1,$ for some $0<\alpha_k\le 1$, then $\pi_k+\pi_1\ge 1$ and therefore Condition (\ref{comproof}) is satisfied.\\
		Then, if $n$ is integer, the condition of Result~\ref{res1} is always fulfilled.
	\end{proof}
	
	

\end{document}